# Evaluación de las técnicas de escalabilidad por calidad para la transmisión de video

*Wilder Castellanos*
*Facultad de Ingeniería. Universidad de San Buenaventura*
*wcastellanos@usbbog.edu.co*

**Resumen**

El incremento significativo del contenido multimedia transmitido sobre internet está demandando nuevas estrategias para asegurar una buena calidad de experiencia por parte de los usuarios. La transmisión de video sobre las redes de datos no es una tarea fácil debido a las diversas variaciones de las condiciones de la red. Una posibilidad de mejorar la calidad de los servicios de video en streaming, es el uso combinado de la codificación escalable de video y los mecanismos cross layer que permiten a las aplicaciones adaptar su tráfico de acuerdo con los recursos disponibles en la red. En este artículo, se presenta una evaluación de las tres principales técnicas de escalabilidad que pueden ser usadas en la codificación de video: CGS (Coarse-Grained Scalability), FGS (Fine Grain Scalability) y MGS (Medium Grain Scalability). Concretamente, se pretende determinar cuál método es más apropiado para la transmisión de video, teniendo en cuenta algunas métricas de calidad como: la relación señal a ruido pico (PSNR, Peak Signal-to-Noise Ratio) y la tasa de frames decodificados. Los resultados obtenidos muestran que la estrategia adaptativa y la escalabilidad MGS ayudan a evitar o a reducir la congestión en la red, obteniendo una mejor calidad en los videos recibidos.

**Abstract**

The significant increase of the transmission of multimedia content over Internet are demanded new delivery strategies to assure a good quality of experience of the users. Transmission of video over packet networks is not an easy task due to multiple fluctuations of the network conditions. One possibility to improve the quality of some video streaming services is the combinate use of the scalable video coding and cross layer mechanisms that allow applications to adapt its traffic stream to the resources network that are available. In this paper, it is presented a performance evaluation of the three main scalability techniques: CGS (Coarse-Grained Scalability), FGS (Fine Grain scalability) and MGS (Medium Grain Scalability). In particular, we focus on determining what method is more appropriated for video transmission taking into account some video quality metrics like PSNR (Peak Signal-to-Noise Ratio) and decoded frame rate. The results reveal that the rate-adaptive strategy and the MGS technique help avoid or reduce the congestion in networks obtaining a better quality in the received videos.

## 1. Introducción

El desarrollo y masificación de los servicios de video en "*streaming*" ha provocado un incremento significativo en la transmisión de contenido multimedia entre las redes de datos. Por ejemplo, de acuerdo al más reciente estudio de Cisco (CISCO Corp., 2017), el tráfico de video desde los dispositivos móviles representará el 78% de todo el tráfico en 2021. También Ericsson pronostica un aumento drástico en el tráfico de video en su informe "*Mobility Report*" (Ericsson, 2016). En dicho estudio se prevé un aumento del tráfico de video desde los dispositivos móviles, pasando de 2.5 Exabytes/mes (cifra de año 2015) a 35 Exabytes/mes en el 2021. Nuevos hábitos en el consumo del contenido audiovisual también están impactando las redes de comunicaciones. Ya que cada vez es más normal ver los contenidos en "streaming"[1], principalmente debido a la masiva explosión del contenido disponible en Internet y de servicios como

---

[1] Término usado para describir la visualización de contenido multimedia a la vez que se está descargando continuamente.

Netflix y Youtube. Pero no solamente los servicios de video en "streaming" llaman la atención de los operadores de red. Con la llegada de las arquitecturas inalámbricas de nueva generación, también se abre a posibilidad de transmitir video de mayor calidad y por lo tanto, la implementación de servicios como IPTV, telecirugía (Mattos & Gondim, 2016) y la televisión inmersiva (Bull, 2014). Otro de los aspectos relevantes de los sistemas de video de próxima generación es el hecho de que la mayoría de estos servicios serán desplegados sobre terminales móviles como *smartphones* y *tablets*. Lo cual supone que la comunicación será inalámbrica. Esto impone la necesidad de implementar mecanismos adaptativos en los servicios de video con el fin de hacer frente a las condiciones variables de las redes inalámbricas. En este contexto cobra importancia la integración de la codificación escalable en el diseño de los sistemas multimedia, ya que dicha técnica permite que los flujos de video se adapten dinámicamente a los recursos disponibles en la red. La codificación escalable de video (H.264/SVC, Scalable Video Coding) (ITU-T, 2013), especifica que un video puede ser codificado en varias capas (una capa base y varias capas de "mejora") que proveen diferentes niveles de calidad. Con este esquema basado en capas se obtienen dos principales ventajas. Primero, que los flujos de video quedan codificados con mayor robustez en entornos con altas pérdidas de paquetes. Segundo, que los videos se pueden adaptar a una red con escasez de recursos (por ejemplo, con enlaces congestionados) por medio de la eliminación o adición de capas. La principal característica de la codificación escalable es el poder configurar un video con tres diferentes tipos de escalabilidad (ver Figura 1): espacial, temporal y de calidad (también llamada escalabilidad SNR- signal-to-noise-ratio).

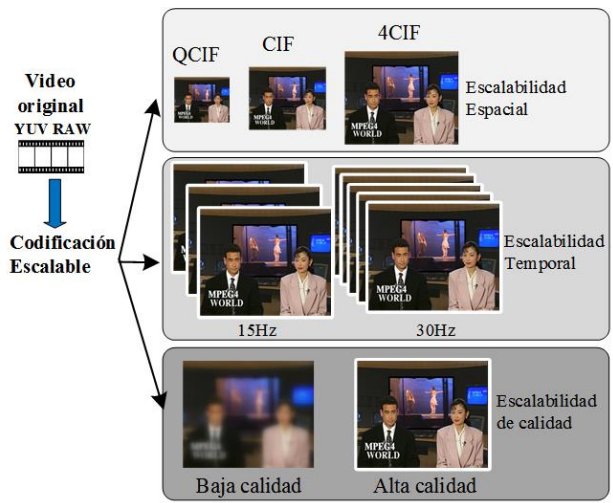

Figura 1: Tipos de escalabilidad en SVC (Castellanos, Guerri, & Arce, 2016)

Tal como se observa en la Figura 1, cuando se usa la escalabilidad temporal, las diferentes capas mejoran la tasa de fotogramas (*frame rate*). Con la escalabilidad espacial la capa base se codifica a una resolución espacial baja y las capas de mejora ofrecen un aumento gradual de la resolución. Finalmente, la escalabilidad en calidad (o SNR) se refiere a escalar en términos del nivel de compresión aplicado al video original. Con la escalabilidad por calidad, la capa base corresponde a una versión muy comprimida de cada fotograma y las capas de mejora proveen más información para incrementar el valor de la relación señal a ruido (esto supone mejorar la calidad del fotograma). Sin embargo, la escalabilidad por calidad del estándar SVC tiene definidas varias técnicas para codificar un video.

En los últimos años el desarrollo de investigaciones sobre SVC ha sido un área muy activa. Por ejemplo, se han implementado varias herramientas software, como las presentadas en las referencias (Castellanos, Guerri, & Arce, 2017) y (Ke, 2012), que permiten desarrollar estudios de simulación sobre plataformas especializadas en la codificación, transmisión y evaluación de videos codificados en SVC. Estas plataformas han facilitado el desarrollo de soluciones para transmitir video sobre diferentes escenarios de red. Por ejemplo, en algunos trabajos como (Hooft et al., 2016) y (Zhang, Liu, & Yuan, 2017) se utiliza la codificación SVC para transmitir video adaptativo sobre el protocolo HTTP. En otros escenarios como las redes vehiculares (VANETs , Vehicular Ad-hoc Networks) y las redes celulares, también se han analizado los mecanismos necesarios para usar la codificación SVC (Feng et al., 2017; Yaacoub, Filali, & Abu-Dayya, 2015). Estos trabajos demuestran la vigencia y la importancia que ha tomado la codificación escalable como una solución que facilita la transmisión de video sobre redes en las cuales las condiciones pueden ser muy adversas. En relación con el análisis de las técnicas de escalabilidad (espacial, temporal o por calidad), algunos autores han implementado algoritmos para mejorar el funcionamiento de dichos esquemas de escalabilidad (Jeong & Song, 2014; Wang, Fang, Jiang, & Chang, 2014). Sin embargo, en lo relacionado con la escalabilidad por calidad no se tienen resultados claros que permitan determinar cuál es el mejor esquema para gestionar este tipo de escalabilidad y su correlación con la calidad de video obtenida.

En este artículo se desarrolla una evaluación de los tres esquemas de escalabilidad por calidad: CGS (Coarse-Grained Scalability), FGS (Fine Grain Scalability) y MGS (Medium Grain Scalability). El propósito de la evaluación fue determinar cuál técnica es más apropiada para la transmisión de video sobre las redes de datos. Esta evaluación se desarrolló mediante la transmisión de videos sobre una red de datos simulada en una plataforma de simulación especializada, en donde se analizaron dos variables: la calidad del video recibido (medida en términos del PSNR, Peak Signal-to-Noise Ratio) y la tasa de frames decodificados.

## 2. La transmisión adaptativa de video

Existen varias técnicas para la transmisión adaptativa de video sobre las redes de comunicaciones. Sin embargo, todas tiene el mismo propósito: adaptar un flujo de video a las condiciones de la red sobre la cual se está transmitiendo. Esto significa que en redes donde la capacidad de transmisión varía continuamente, se podrían ir ajustando algunos los parámetros del video que se está transmitiendo, con el fin de adaptarlo a dichas variaciones. Esto permitiría, por una parte, mejorar la calidad de los videos que se reciben al final de la transmisión y, por otra parte, ayudar al control de la congestión de paquetes.

Una de las técnicas más promisorias para adaptar un video a las condiciones de la red, es mediante la utilización de la codificación escalable de video (también conocida como SVC, Scalable Video Coding). Esta codificación convierte un video en un flujo de datos compuesto por varias "capas". Hay una capa base que contiene la representación más básica del video, es decir la representación de más baja calidad. También se generan otras capas, denominadas capas de mejora, que pueden adicionarse a la capa base para incrementar la calidad del video (por ejemplo, aumentando el frame rate o la resolución espacial). Este esquema por capas permite que el emisor de un video pueda adaptar su tasa de envío de datos (bit rate), añadiendo o eliminando capas del video. Esta adición o sustracción de capas puede hacerse a medida que las condiciones de la red van variando, por ejemplo, cuando se presentan fluctuaciones en el ancho de banda disponible (Ver Figura 2).

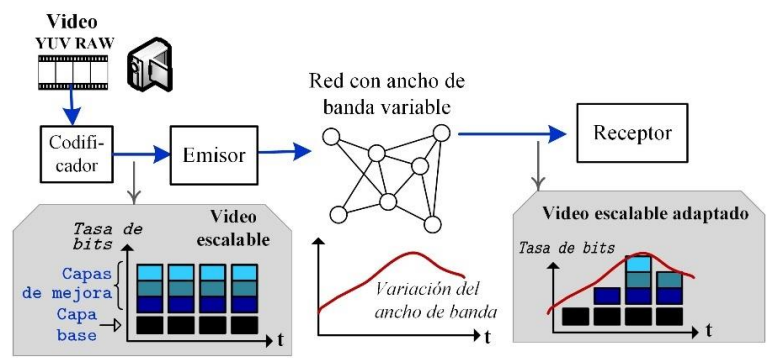

Figura 2: Transmisión adaptativa de video SVC

En la codificación SVC existen tres tipos de escalabilidad: espacial, temporal y de calidad (o escalabilidad SNR). La escalabilidad espacial permite codificar la capa base a una resolución[2] espacial baja y las capas de mejora ofrecen un incremento gradual de la resolución. Por lo tanto, el uso de la escalabilidad espacial permite decodificar el video con una menor resolución que la original, decodificando solo la capa espacial más baja y descartando las otras capas espaciales. Cuando se usa la escalabilidad temporal, las capas mejoran la tasa de imágenes por segundo (frame rate). La escalabilidad por calidad consiste en generar una capa base con un alto nivel de compresión y capas de mejora que gradualmente van introduciendo más información con el fin de incrementar la relación señal a ruido de cada fotograma (SNR, Signal-Noise-Ratio).

---

[2] Entiéndase por resolución como el tamaño en pixeles del fotograma

De acuerdo con el estándar H.264/SVC, es posible usar una escalabilidad combinada, utilizando más de un tipo de escalabilidad. Por ejemplo, un video puede tener escalabilidad temporal y de calidad al mismo tiempo. Tal es el caso de la estructura que se muestra en la Figura 3, donde un video codificado en SVC tiene tres niveles temporales {T0, T1 y T2} y dos niveles de calidad {Q0 y Q1} (en este ejemplo no se usa la escalabilidad espacial). Esta estructura sugiere que, si transmitimos el nivel T0, los únicos fotogramas que se enviarían serían los fotogramas 0 y 4. Y, a medida que se transmitan los otros niveles temporales, se van integrando al flujo de imágenes los demás fotogramas. Por consiguiente, se aumentaría el frame rate del video. Adicionalmente, tenemos dos versiones de cada fotograma, una versión de baja calidad (Q0) y otra de alta calidad (Q1). En consecuencia, se obtienen en total 6 capas que son el resultado combinar los niveles temporales y de calidad. La capa base, en esta estructura estaría integrada por el nivel temporal de más bajo frame rate (T0) y del nivel de calidad más bajo (Q0). De igual manera, la capa de mejora más alta (capa 5) estaría conformada por el nivel temporal T2 y el nivel de calidad Q1. Cada una de las 6 capas tiene tres parámetros que la identifican: DID, TID y QID (también son conocidos como parámetros DTQ): donde DID es el identificador del nivel espacial, TID es el identificador del nivel temporal y QID es el identificador del nivel de calidad. Además, en este ejemplo se muestran las dependencias que existen entre los fotogramas que conforman el video. Estas dependencias se representan por medio de las flechas que unen los diferentes fotogramas. Por ejemplo, el fotograma 1 de la capa T2Q0 depende de la capa T0Q1 y también de la capa T1Q1. Debido a estas dependencias, descartar una capa de un fotograma (tal como el fotograma 4) puede afectar la calidad de los fotogramas dependientes (e.g. los fotogramas 3 y 5)

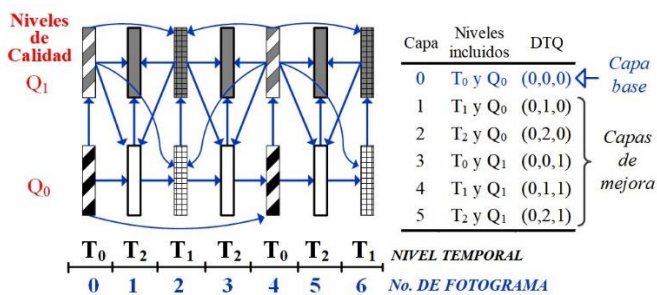

Figura 3: Estructura de un video codificado con escalabilidad temporal y de calidad (Castellanos et al., 2017)

Para obtener el video en formato SVC es necesario utilizar un codificador H.264/SVC, por ejemplo el codificador de referencia JSVM (Joint Scalable Video Model) (Moving Pictures Experts Group, 2011). Los codificadores H.264 tiene una estructura de dos sub-sistemas (ver Figura 4): el subsistema de codificación de video (VCL, Video Coding Layer) y el subsistema de adecuación a la red (NAL, Network Abstraction Layer).

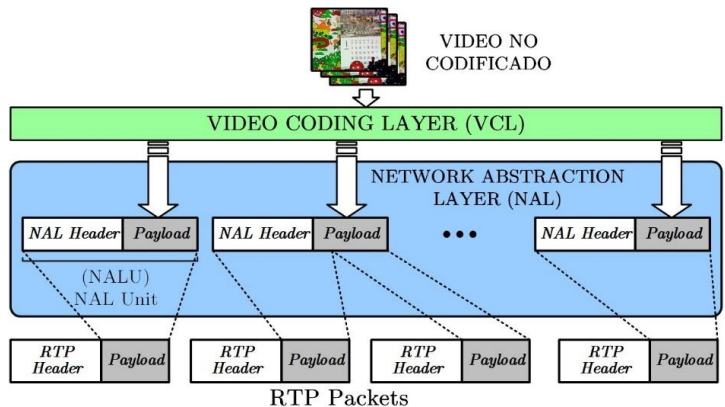

Figura 4: Estructura del codificador H.264/SVC

Básicamente, el subsistema VCL se encarga de la codificación del video y el subsistema NAL agrupa los bits resultantes de la codificación en unidades que puedan ser transportadas a través de una red de datos. Estas unidades conformadas en el subsistema VCL se llaman NALUs (NAL Units). Cada NALU consiste en un número entero de bytes del video codificado. Posteriormente, se conforman paquetes de capa de transporte (por ejemplo, aplicando el formato del protocolo RTP, Real-Time Transport Protocol) con los bits de las NALUs. Tal como se observa en la Figura 4, el tamaño de un paquete RTP puede variar dependiendo del número de NALUs que pueda transportar. También depende del tamaño de las NALUs, ya que aquellas NALUs que contienen la información de los fotogramas de la capa base tendrán más bytes. Cada NALU se identifica con los parámetros DTQ que permite identificar a que capa pertenece una determinada NALU. Esta es una característica importante, ya que de esta manera una NALU (por ejemplo, una que lleva bits de una capa de mejora) puede ser identificada y extraída del flujo de video SVC para reducir la tasa de bits.

## 2.1. La escalabilidad por calidad en SVC

El estándar H.264/SVC provee tres esquemas de escalabilidad por calidad: escalabilidad por granularidad gruesa (CGS, Coarse Grain Scalability), escalabilidad por granularidad media (MGS, Medium Grain Scalability) y escalabilidad por granularidad fina (FGS, Fine Grain Scalability) (Schwarz, Marpe, & Wiegand, 2007).

En el esquema CGS cada capa de mejora es estimada por separado, esto es importante ya que no existen dependencia entre las capas de mejora y la capa base. Por lo tanto, pérdidas de información de la capa base no influye en la estimación de las capas de mejora. Otra ventaja adicional es que este esquema tiene un menor costo computacional. Sin embargo, se tienen algunas desventajas, por ejemplo, si una parte de la capa de mejora se pierde entonces se obtiene un error en la estimación de la secuencia posterior (ver Figura 5a).

Con el esquema de FGS cada capa de mejora se predice a partir de la capa base (Figura 5b). Con ello se evita que, si se pierden los paquetes correspondientes a la capa de mejora durante unos instantes de tiempo, se pueda continuar decodificando la capa de mejora a partir de la capa base. Sin embargo, tenemos una eficiencia de codificación muy baja, ya que solamente los fotogramas de la capa base son usados para predecir las capas de mejora y, además, la capacidad de procesamiento invertida para codificar estas capas de mejora no se puede aprovechar para codificar los siguientes fotogramas.

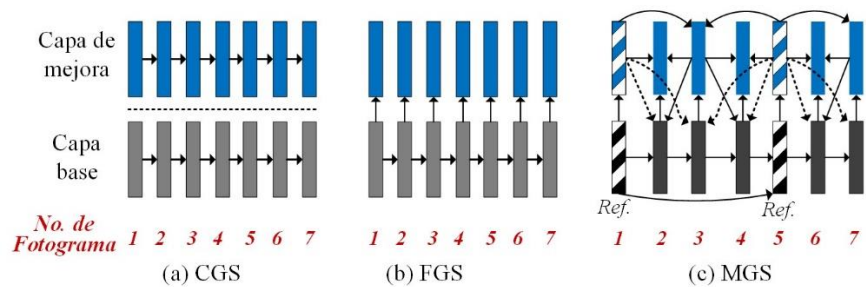

Figura 5: Métodos de escalabilidad de calidad

El esquema MGS (Figura 5c) aumenta la complejidad de la codificación, pero ofrece más flexibilidad ya que permite elegir la capa de mejora que se va a usar para la predicción de los siguientes fotogramas. En MGS es necesario definir unos *fotogramas de referencia,* los cuales sirven para hacer actualizaciones periódicas en la capa base. Por ejemplo, los fotogramas 1 y 5 son los fotogramas de referencia y a partir de su capa base se predice la capa de mejora. A su vez, la capa de mejora del fotograma de referencia puede ser usada para predecir la capa base y la capa e mejora del fotograma 6 y 7. Pero además permite predecir las capas del fotograma 3 y 4. En la referencia (Schwarz et al., 2007) se puede consultar información más detallada sobre la técnica de codificación de video escalable y los diferentes esquemas de escalabilidad.

## 3. Evaluación de las técnicas de escalabilidad por calidad

Para poder evaluar los tres esquemas de escalabilidad (CGS, FGS y MGS) fue necesario utilizar una plataforma de simulación de redes que además permitiese configurar tráfico de video. La plataforma seleccionada fue SVCEval-RA (SVC Evaluation platform for Rate-Adaptive Video) (Castellanos et al., 2017). Este software contiene las herramientas necesarias para codificar un flujo video en SVC y transmitirlo por una red de comunicaciones. Esto último es posible gracias a la integración que existe entre SVCEval-RA y el simulador de redes NS-2. Además, esta integración permite simular la transmisión del video SVC sobre diversas tecnologías de red tanto cableadas como inalámbricas y usando diversos protocolos de red. Al tener a disposición todos los recursos de NS-2, también es posible evaluar diferentes métricas relacionadas con el funcionamiento de una red de datos, tales como el retardo, la tasa de pérdidas de paquetes, entre otras. Uno de los aspectos esenciales en la simulación de transmisión de video SVC, es la capacidad de replicar la adaptación del flujo SVC añadiendo o eliminando capas de mejora, de acuerdo con la variación del ancho de banda. Esta capacidad es otorgada por los algoritmos implementados en SVEval-RA. En la Figura 6 se muestra un diagrama de flujo de la herramienta de simulación SVCEval-RA y de la metodología utilizada para configurar una simulación. Principalmente consiste en tres principales fases: el pre-procesamiento, la simulación y el pos-procesamiento.

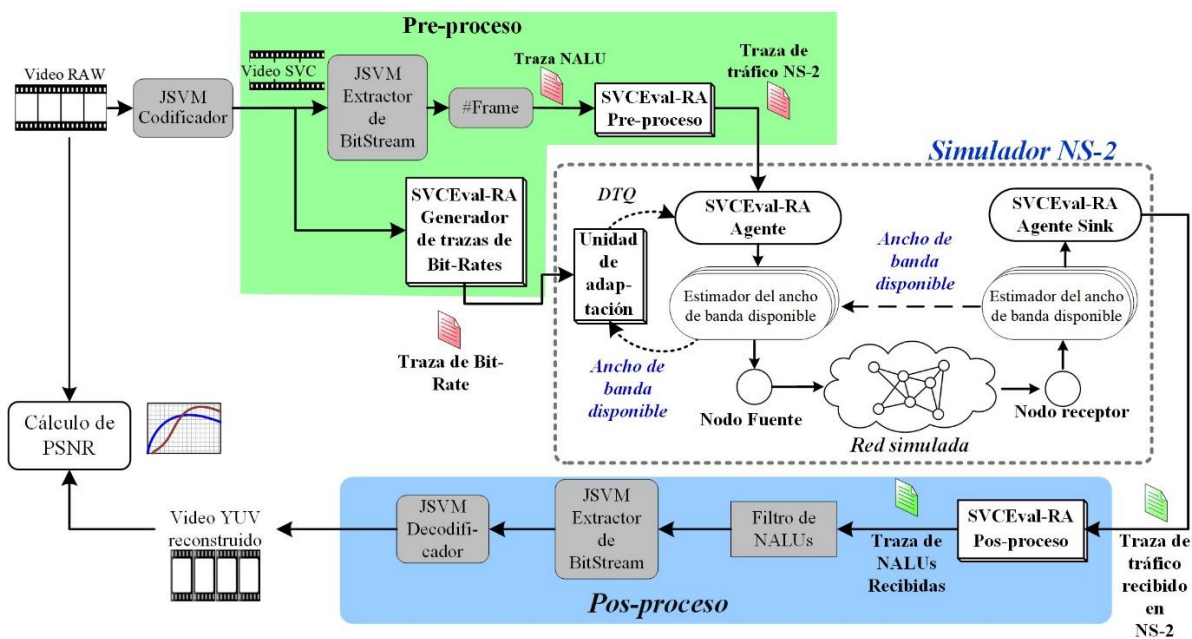

Figura 6: Metodología para la implementación de una simulación en SVCEval-RA

En la fase de pre-procesamiento, el video original en formato YUV es codificado en formato SVC, estableciendo el tipo escalabilidad y el número de capas utilizando del software JSVM (Joint Scalable Video Model) (Moving Pictures Experts Group, 2011). Posteriormente, por medio de la herramienta *BitStreamExtractor*, incluida en JSVM, se genera un fichero de texto ("traza de NALUs") con un listado de todos los paquetes en los que se ha segmentado la información de cada NALU. Para cada paquete se almacena los siguientes datos: el tamaño del paquete, los valores DTQ de la NALU y el tipo de paquete. Previo a la fase de simulación, se generan dos trazas adicionales, la cuales son generadas a través del software SVCEval-RA. La primera traza es la "traza de bit-rates", la cual contiene información acerca de la tasa de transmisión requerida para transmitir cada capa SVC. Particularmente, los datos que contiene esta traza son: el identificador de capa, la tasa de fotogramas (frames por segundo), la tasa de transmisión (Kbps), los parámetros DTQ y el PSNR (Peak Signal-to-Noise Ratio). La segunda traza es la "traza de tráfico NS-2", la cual es una adaptación de la traza de NALUs al formato aceptado por NS-2.

En la fase de simulación, NS2 utiliza la información incluida en la "traza de tráfico NS-2" para generar los paquetes que van a conformar el flujo de video y los transmite por la red simulada. Como resultado de la simulación, se obtiene el archivo "traza de tráfico recibido", el cual contiene un registro por cada paquete, similar al fichero de entrada, pero adicionalmente se indica el retardo experimentado por cada paquete durante su transmisión. En caso de que un paquete se elimine durante la transmisión, el correspondiente registro no aparecerá en el fichero de salida. Durante la transmisión de los paquetes, la red periódicamente hace la estimación del ancho de banda disponible, teniendo en cuenta todo el tráfico que circula por la red, y envía dicha estimación a la unidad de adaptación. Tal como se muestra en la Figura 6 , en dicho módulo se determinan los parámetros DTQ de la capa más alta que puede transmitirse sin superar la limitación del ancho de banda. Por lo tanto, la capa seleccionada es aquella con un bit-rate menor o igual que la tasa de transmisión soportada por la red. A partir de los parámetros DTQ, el agente de software SVCEval-RA integrado en el simulador, filtra los paquetes correspondientes a dicha capa (y sus capas inferiores), para que solo estos sean transmitidos por la red móvil ad hoc.

Al finalizar la simulación el archivo "traza de tráfico recibido" es procesado por SVCEval-RA para obtener la "traza de NALUs recibidas". Esta última traza necesita ser filtrada con el fin de eliminar las NALUs que han superado el límite de retardo. Además, es necesario reordenar las NALUs de acuerdo con el orden de envío y, por último, eliminar las NALUs que no tienen satisfechas sus dependencias para la decodificación. Después de filtrar la "traza de NALUs recibidas", y con ayuda de la herramienta *BitStreamExtractor*, se reconstruye el video SVC, el cual a continuación se decodifica a formato YUV. Finalmente se calcula el PSNR, comparando el video original en formato YUV y video reconstruido con el fin de medir la calidad del video recibido. Adicionalmente, otras métricas pueden ser calculadas a partir de la traza de tráfico NS-2, tal como la tasa de paquetes perdidos y el retardo.

## 4. Evaluación

En esta sección se presentan los parámetros configurados durante las simulaciones, así como las características de los videos utilizados y la codificación aplicada. El video de prueba utilizado para las simulaciones se generó concatenando cuatro veces el video "MOBILE". Este video hace parte de la galería de videos de Xiph.org Foundation («Xiph.org Video Test Media», 2016). El video concatenado fue codificado en SVC tres veces, generando tres versiones diferentes del video. Cada una de las tres versiones tiene un esquema diferente de escalabilidad de calidad. El códec SVC utilizado fue el códec JSVM (Moving Pictures Experts Group, 2011). En la Figura 7 se muestra de forma gráfica el proceso de codificación del video y se describen otros parámetros relacionados generación de los videos SVC.

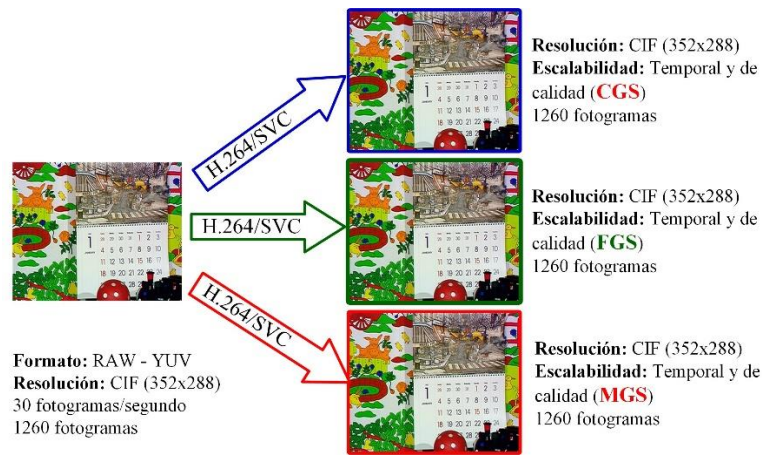

Figura 7: Generación de los videos de prueba

### 4.1. Entorno de simulación

En la Figura 8 se describe el escenario de simulación utilizado: tanto la topología de la red (Figura 8a), como la variación del ancho de banda disponible en este escenario (Figura 8b). De acuerdo con dichas figuras, la fuente de video (nodo S1) y el destino del video (D1) se ubican en los extremos de la red. La transmisión del video inicia en el instante t=10s e inicialmente es el único tráfico presente en la red. De esta manera, el ancho de banda disponible, desde un extremo hasta el otro, es de 1.5Mbps. A medida que avanza la simulación se van incorporando gradualmente otros tráficos adicionales: primero se adiciona el tráfico entre A y D, luego entre B y C, y finalmente entre E y F. Esto provoca una continua variación del ancho de banda disponible que percibe el flujo de video entre S1 y D1, ya que se producirá un cuello de botella en el enlace entre R4 y R5. Esta variación se induce con el fin de evaluar la capacidad de adaptación del flujo de video a las variaciones de las condiciones de la red. La variación del ancho de banda disponible se describe en la Figura 8b, donde se identifican dos partes: la primera parte corresponde al intervalo de tiempo entre 10s y 42s, en donde el ancho de banda disponible de la red decrece en intervalos de 4 segundos desde 1.5Mbps a 0.2Mbps. En la segunda parte, el ancho de banda se incrementa desde 0.2Mbps hasta 1.5Mbps en intervalos de 4 segundos. Este escenario permite evaluar el comportamiento del flujo de video ante unas abruptas variaciones del ancho de banda. Estas variaciones son comunes en la comunicación de datos en Internet, por ejemplo, cuando se generan cuellos de botellas en ciertos enlaces debido al tráfico de otros usuarios (Akhshabi, Begen, & Dovrolis, 2011). Esto quiere decir que la variación del ancho de banda utilizada en este estudio no es exclusiva de la topología de red de la Figura 8a, sino que se presenta en muchos otros escenarios reales.

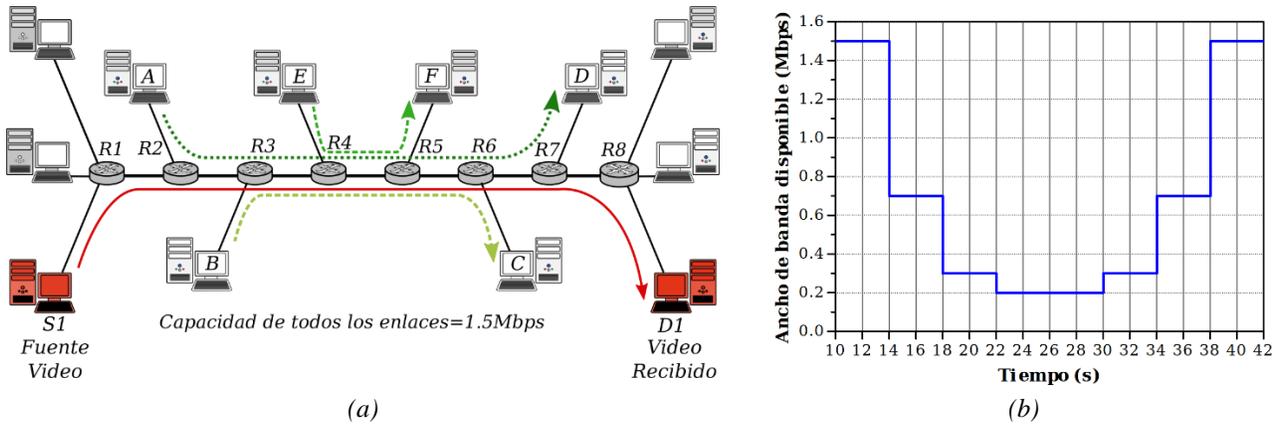

Figura 8: Escenario de evaluación: (a) red simulada (b) variación del ancho de banda disponible entre S1 y D1

Cada uno de los tres videos codificados en SVC (CGS, FGS y MGS) fueron transmitidos sobre el escenario de red descrito anteriormente, utilizando el simulador NS2. Al final de cada simulación se obtuvieron los videos recibidos, los cuales se compraron entre sí para determinar cuál sufrió mayor degradación.

### 4.2. Parámetros de evaluación

En el escenario simulado se calcularon dos métricas. La primera métrica consistió en cuantificar la calidad de los videos recibidos. Esto se determina calculando la relación señal a ruido pico de la luminancia (conocida como Y-PSNR, Peak Signal-to-Noise Ratio). El cálculo de la PSNR nos ofrece una cuantificación clara de la distorsión que sufre el video con un costo computacional bajo. Esta métrica es una de las más utilizadas para evaluar la calidad de un vídeo después de su transmisión por una red. La Ec. (1) muestra la definición de la PSNR.

$$PSNR = 10 \cdot Log_{10}\left(\frac{255^2}{MSE}\right) \quad donde,$$
$$MSE = \frac{1}{NM}\sum_{m=0}^{M-1}\sum_{n=0}^{N-1}\left|I_{org}(m,n) - I_{rec}(m,n)\right|^2$$

(1)

donde $I_{org}$ es la imagen original e $I_{rec}$ es la imagen recibida después de haber sido transmitida por una red; *M, N* es el tamaño de la imagen y *MSE* es el error cuadrático medio. De acuerdo a la Ec. (1), la PSNR es la comparación, pixel a pixel, de cada uno de los fotogramas que conforman los dos videos. De esta manera, se obtiene un valor numérico que refleja la distorsión sufrida por el video recibido con respecto al original.

La segunda métrica calculada fue el porcentaje de fotogramas decodificables, el cual se define como el número de fotogramas decodificables sobre el número total de fotogramas enviados por la fuente de vídeo (Kao, Ke, & Shieh, 2006). Por definición, un fotograma es considerado decodificable sólo cuando se recibe correctamente una cantidad de información de codificación que supere un cierto umbral de decodificación. Por consiguiente, un fotograma se considera decodificable siempre y cuando todos los paquetes de ese fotograma y los paquetes de los que ese fotograma depende, son correctamente recibidos.

## 5. Resultados

A continuación, se presentan los resultados obtenidos al final de las simulaciones. En la Figura 9, se presenta la variación de la calidad de los videos (Y-PSNR) en función del tiempo de simulación. Estos resultados muestran que los videos MGS y FGS fueron adaptando sus flujos añadiendo y eliminando capas de mejora a medida que el ancho de banda fue variando. En cuanto al video CGS, logró adaptarse a la variación decreciente del ancho de banda hasta cuando el cuello de botella se volvió muy restrictivo. En dicho intervalo (desde 22s hasta los 30s) se produjeron muchas pérdidas de paquetes que afectaron la correcta decodificación de las capas de mejora. Estas pérdidas también provocaron que los fotogramas posteriores no se lograran decodificar, debido a la propagación de las pérdidas. En consecuencia, se obtuvo una mayor degradación del video recibido.

La Figura 10, muestra la PSNR media en cada segmento, lo cual permite observar con mayor claridad la calidad de los videos recibidos en cada uno de los intervalos de variación del ancho de banda. Se destaca que el video con mayor calidad en todos los segmentos fue el video MGS. En promedio al final de la simulación, este video obtuvo un PSNR de 27.4dB, el video FGS de 23.4dB, mientras que el video CGS obtuvo 20.3dB. Esto significa una diferencia de 4 dB en la calidad del video MGS con respecto al video FGS y de 7dB con relación a CGS, lo cual se refleja en una mayor calidad percibida por el usuario. Además, la mayor diferencia entre MGS y FGS se obtuvo en intervalo de tiempo en el cual el ancho de banda era más restrictivo (22s-30s). Esto se debe a la pérdida de información de la capa base del video FGS, lo cual conlleva a que, tanto los fotogramas de la capa base como los de las capas de mejora, no se puedan decodificar correctamente ya que los errores de decodificación se propagan a lo largo del video. Comparando las codificaciones FGS y CGS se evidencia que el esquema FGS tiene una mejor capacidad de recuperarse una vez se empiezan nuevamente a aceptar capas de mejora. Estos resultados permiten inferir que la codificación MGS es más robusta, lo cual la hace más apropiada para ser utilizada en escenarios de altas pérdidas, tal como pueden ser las redes móviles.

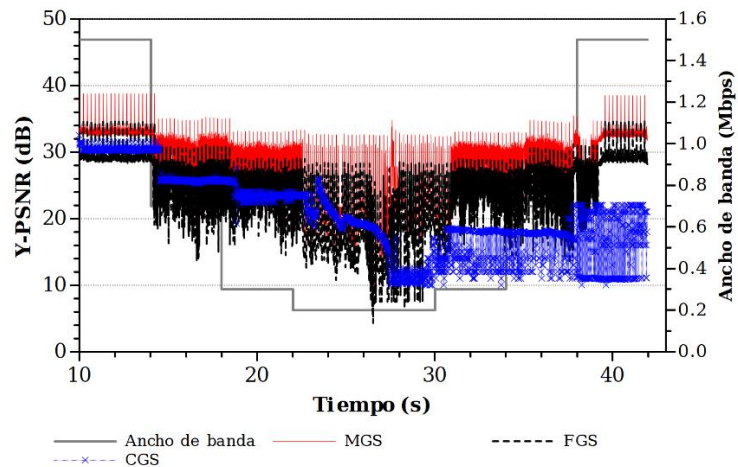

Figura 9: Variación instantánea de la PSNR

Con el fin de proveer una valoración subjetiva de la calidad de experiencia (QoE, Quality of Experience) del usuario, se utilizó la escala MOS (Mean Opinion Score). En términos generales, la escala MOS es un indicador numérico de la calidad percibida por el usuario después de la transmisión y decodificación de un video. Debido a que la escala MOS es una escala subjetiva, su evaluación requiere una interpretación humana. Sin embargo, esto requiere de usualmente se puede lograr una aproximación a partir de la estimación de una métrica objetiva, tal como la PSNR. Para este caso, se asumió el mapeo planteado en la Tabla 1, el cual facilita la conversión de PSNR a MOS. De acuerdo a esta aproximación, la QoE del video MGS sería de *Regular*, lo cual implicaría para el usuario una distorsión *Tolerable*, mientras que los videos FGS y CGS tendría un MOS de 2 (*Pobre*), es decir una distorsión que resultaría molesta para el usuario.

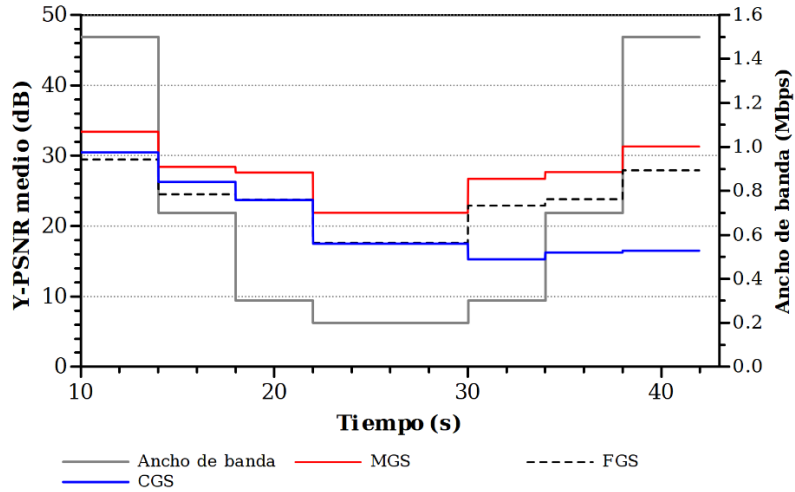

Figura 10: PSNR promedio vs. Tiempo de simulación

Tabla 1: Tabla de conversión PSNR – MOS (Klaue, Rathke, & Wolisz, 2003)

| PSNR [dB] | Escala MOS | Distorsión |
| --- | --- | --- |
| >37 | 5 (Excelente) | Imperceptible |
| 31–37 | 4 (Bueno) | Perceptible, pero tolerable |
| 25–31 | 3 (Regular) | Tolerable |
| 20–25 | 2 (Pobre) | Molesta |
| <20 | 1 (Malo) | Muy molesta |

En la Figura 11 se muestra el porcentaje de fotogramas que se decodificaron con cada esquema. También con esta métrica se identifica un mejor rendimiento de la codificación MGS. Con este esquema se logra la mayor tasa de fotogramas decodificables. Alrededor del 72% de los fotogramas fueron decodificados correctamente, mientras que con FGS se decodificaron el 61% y con CGS solo el 45%.

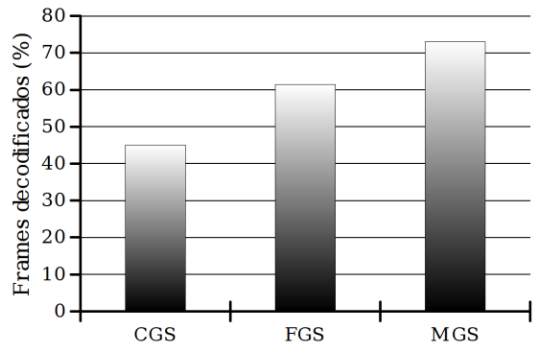

Figura 11: Porcentaje de frames decodificados

## 6. Conclusiones

En este artículo se evaluaron las tres principales técnicas de escalabilidad en calidad que se pueden usar cuando se codifica un video en H.264/SVC. El principal propósito fue encontrar el mejor esquema para este tipo de escalabilidad y su correlación con la calidad de video obtenida. La evaluación de dichos esquemas se realizó mediante la transmisión flujos de video SVC a través de un escenario de red simulado en el cual el ancho de banda disponible variaba por intervalos de tiempo de 4 segundos. Las simulaciones se ejecutaron sobre una plataforma especializada en este tipo de estudios.

Los resultados obtenidos muestran en primer lugar que, independientemente del esquema usado para la escalabilidad en calidad, la codificación SVC le permite al flujo de video ajustarse a las condiciones de la red. Los resultados también demuestran que la escalabilidad MGS ofrece una robustez adicional, en comparación a FGS y CGS. Especialmente en escenarios donde las condiciones restrictivas de la red hacen que las pérdidas de paquetes aumenten y, por consiguiente, sea más difícil la decodificación de los fotogramas del video. Esto se debe a que este tipo de escalabilidad incrementa la eficiencia de codificación debido a su esquema basado en fotogramas de referencia. Este hecho trae varias ventajas: primero que todo, permite que los errores de decodificación no se propaguen a lo largo de todo el video. Esto quiere decir que si algún fotograma no puede ser decodificado algunos fotogramas dependientes no podrán ser decodificados. Sin embargo, este hecho será subsanado cuando sea decodificado un fotograma de referencia. Por otra parte, el esquema basado en fotogramas de referencia permite que se pueda cambiar de capa en cualquier momento de la decodificación.